\documentclass[twocolumn]{elsart3}
\usepackage{graphicx}
\usepackage{amssymb}

\begin{document}

\begin{frontmatter}

\title{Fabrication and characterization of scanning tunneling microscopy superconducting Nb tips having
highly enhanced critical fields}
 \author[unisa]{A. Kohen\corauthref{cor1}},
  \ead{kohen@gps.jussieu.fr}
 \author[unisa]{Y. Noat},
 \author[unisa]{T. Proslier},
 \author[unisa]{E Lacaze},
 \author[swiss]{M. Aprili},
 \author[unisa]{W. Sacks}
 and
 \author[unisa]{D. Roditchev}
\corauth[cor1]{Corresponding author.}
 \address[unisa]{Groupe de Physique des Solides, UMR7588 au CNRS, Univ. Paris 6 and Paris 7, 140 rue de Lourmel, 75015 Paris France}
 \address[swiss]{Laboratoire de Spectroscopie en Lumi\`{e}re Polaris\'{e}e-CNRS
ESPCI 10, rue Vauquelin 75005 Paris }



%\title{Fabrication and characterization of scanning tunnelling microscopy superconducting Nb tips with highly enhanced critical fields}


\author{}

\address{}

\begin{abstract}
We report a simple method for the fabrication of Niobium
superconducting (SC) tips for scanning tunnelling microscopy which
allow atomic resolution. The tips, formed in-situ by the
mechanical breaking of a niobium wire, reveal a clear SC gap of
1.5 meV and a critical temperature T$_c$\,=\,9.2$\pm$0.3 K, as
deduced from Superconductor Insulator Normal metal (NIS) and
Superconductor Insulator Superconductor (SIS) spectra. These match
the values of bulk Nb samples. We systematically find an enhanced
value of the critical magnetic field in which superconductivity in
the tip is destroyed (around 1\,T for some tips) up to five times
larger than the critical field of bulk Nb (0.21\,T). Such
enhancement is attributed to a size effect at the tip apex.
\end{abstract}

\end{frontmatter}

% main text
\section{Introduction}
\label{Intro}
Among the variety of surface techniques, Scanning
tunneling microscopy and spectroscopy (STM STS) has a special role
in the investigation of surface topography and local electronic
properties on the atomic scale \cite{Binning}. The device is based
on the precise measurement of the vacuum tunneling current between
an atomically sharp tip and a sample as a function of the applied
voltage. The magnitude of the current depends on the tip sample
distance and on the density of electronic states of both the tip
and the sample. Usually the STM tip is made of a normal metal in
which the density of electronic states (DOS) near the Fermi level can be
considered, in a first approximation, constant. Thus the measured
current as a function of tip position reflects changes in either
the sample DOS or the topography.

Meservey\cite{Meservey} had suggested the use of an SC material
for the fabrication of the STM tip. One advantage of an SC tip is
the enhanced spectroscopic resolution due to the singularity in
the SC DOS at the gap edge. This reduces significantly the thermal
smearing, due to the Fermi-Dirac distribution, in comparison to
the case of a normal metal tip. Of major interest, an SC tip with
an SC sample allows, in principle, direct cooper pair tunneling
(Josephson current\cite{Josephson}), which is well known in point
contact and planar SIS junctions\cite{Jaklevic}. In the
investigation of local electronic properties of superconductors,
the STM is almost exclusively used with a normal tip, thus
exploiting single electron processes, resulting in the well known
BCS tunneling DOS. The gap function is inferred from the spectrum
but does not indicate directly the existence of a condensate. With
an SC tip on the other hand the appearance of a Josephson current
is a definite signature of a coherent state.

In general a proper STM tip has to be sharp enough
to allow for good spatial resolution (in the best case atomic
resolution). From a theoretical point of view it was questioned
whether an atomically sharp tip would still show SC properties.
One certainly expects that the geometry and dimensions of the tip
could have an effect on it's properties as, for example, in the case of
an applied magnetic field. Additionally, one must avoid the
oxidation of the surface layer of the tip, which for normal metals
is easily achieved, for example, by mechanically cutting a PtIr wire in
air. For an SC tip this is not so trivial: SC elements
from which tips can be formed like Al, Nb and Pb oxidize rapidly.
Other materials, such as high temperature superconductors, suffer from
surface degradation and the difficulty in controlling the
geometry, due to their complex structure and composition.

Pan and Hudson\cite{Pan} have shown that atomic resolution and
superconductivity can be obtained at the same time. They have used
a mechanically cut Nb wire to obtain a sharp tip and further apply
a voltage pulse between the tip and an Au target inside the STM
vacuum chamber. This removes the oxide and further sharpens the
tip. The tips were indeed superconducting and allowed for atomic resolution;
however, the measured gap values varied from one tip to the other,
ranging from a few tenths of a meV and up to 1.5 meV.  The authors
suggest that the superconductivity measured at the end of the tip
is in fact due to a proximity effect induced by the bulk. In such
a scenario, the amplitude of the measured gap can vary due to
changes in tip apex geometry and composition. The latter are
claimed to be a result of the voltage pulse application.

Naaman et al.\cite{Naaman}have used a two layer deposition method
in which a $5000\,$\AA\, layer of Pb is deposited on a
mechanically cut PtIr tip and then covered by a $30\,$\AA\, layer
of Ag. In this method, the Ag layer serves as a protection against
the oxidation of the lead. The superconductivity of the tip is
based on the proximity effect and a gap is induced in the Ag
layer. Finally Suderow et al.\cite{Suderow} have used a method in
which the STM tip is successively driven into and pulled out of a
Pb layer and results in the formation of a Pb tip. As the process
is done inside a vacuum chamber and at a temperature of $\sim
4.2\,$K, the resulting tip is mechanically stable and does not
oxidize.

The last two methods are inferior to the first one, as they use Pb which has a
lower critical temperature and gap value ($T_{c}=7.2\, $ K
; $\Delta(T=0)=1.3\,$ meV) in comparison to Nb, being the element
with the highest critical temperature ($T_{c}=9.2\,$K; $\Delta
(T=0)=1.5\,$ meV). This limits the temperature range in which the tip
can be used and enlarges the effect of thermal fluctuations on the
measured Josephson current \cite{Smakov}. Giubileo et
al.\cite{Giubileo} have used a MgB$_{2}$ grain glued to a PtIr tip
as a STM tip, this technique allowed obtaining atomic resolution
and has the advantage of a high critical temperature (~39K) and a
relatively large SC gap ($2-7\,$ meV). However spatial resolution
is not easily reproduced in tips of this kind and the non trivial
two band effects in MgB$_{2}$\cite{Giubileo} hinder the
interpretation of the measured conductance spectra.

Here we
propose a new and simple method for the preparation of SC Nb STM
tips which does not require a voltage pulse application. The
method is based on mechanically cutting an Nb wire inside the
STM's vacuum chamber (a similar method was used to create a
ferromagnetic tip by Koltun et al.\cite{Koltun}). The resulting Nb
tips allow obtaining atomic resolution topographic images and have
a reproducibly measured gap value of $1.5\,$ meV which matches the
value obtained for bulk Nb samples.  By following the temperature
dependence of the conductance spectra measured using an Nb tip and
an Au sample we found the SC gap to close at a temperature of
$9.2\pm0.3\,$ K which coincides with the critical temperature of
bulk Nb samples.  A clear deviation from the bulk properties is
observed under the application of magnetic field. All tips exhibit
a strongly enhanced value of the critical field with a maximum
measured value of $\sim1\,$ T, $5$ times larger than the bulk value
of $0.2\,$ T. We explain this enhancement as a result of the
confinement of the superconducting phase by the magnetic field to
a nano-metric region at the tip apex.

% \appendix

\section{Tip preparation and Samples}
\label{tip} The tips are made from a niobium wire of $0.25\,$mm
diameter from which we cut a $\sim 4\,$ cm piece. Using a surgical
knife we then create a thin groove around the wire thus creating a
weak point. Following, one end of the wire is glued using an ultra
high vacuum compatible conducting silver epoxy to a tip holder and
the other end is tied to create a loop as shown in figure 1a. The
tip and tip holder are then put into an autoclave oven for a 1
hour baking of the glue at a temperature of $100\,^o$C. The final
cutting of the tip is done inside the vacuum chamber which
contains the home-made low temperature (4.2K) STM at a vacuum of
$\sim 4\cdot10^{-8}$ mbar and at ambient temperature. The cutting
process is done using a mechanical manipulator which allows us to
first place the loop around a metallic pin and subsequently to
pull the tip holder. This stretches the Nb wire until it breaks at
the designated weak point thus creating the tip (see fig 1b).  We
then install the tip in the STM head, perform the coarse approach
and lower the microscope to the bottom of a cold finger where,
owing to cryo-pumping, the vacuum is  much better than in the
upper part of the chamber. This entire process from the moment the
tip is cut lasts around $3$ minutes and is short enough to avoid
the creation of an oxide layer on the tip apex. Longer preparation
times, or poorer vacuum conditions, increase the number of tips
showing insulating behavior at low temperature, indicating the
formation of an oxide layer at the tip apex. Under proper
conditions, roughly $75$ percent of the tips show superconducting
properties at low temperatures.

\begin{figure}
\centering
  \includegraphics[angle=-90, width=7cm]{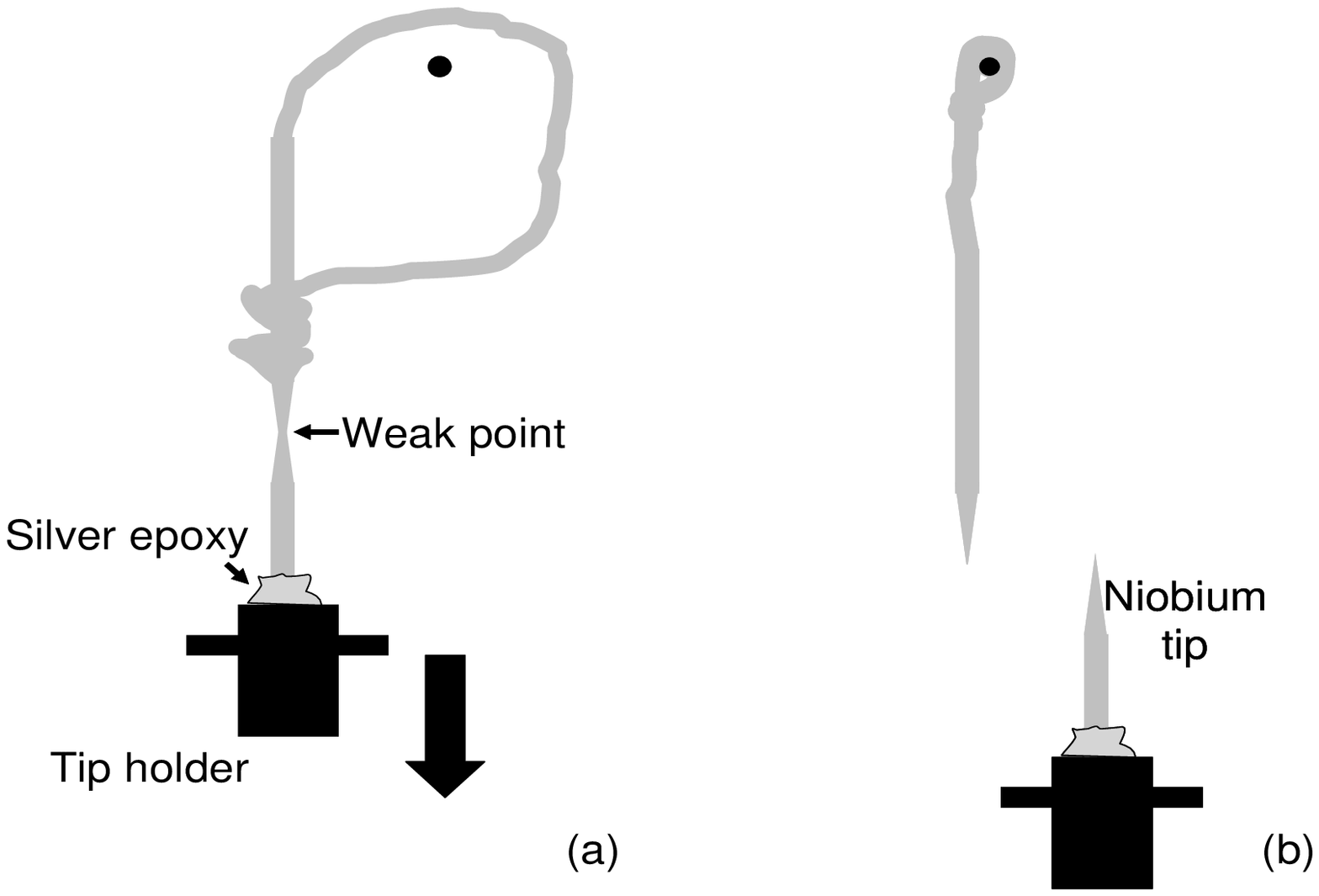}\\
  \caption{Schematic drawing of the tip preparation procedure:
  a) Nb wire prior to it's breaking. Pulling direction by the mechanical
  manipulator is shown by an arrow
  b) Nb tip and the remaining Nb wire part after the break.}
  \label{tip}
\end{figure}

To characterize the tip properties we have used three kinds of
samples:  Au samples of $2000$ \AA\, in thicknesses deposited on
Mica substrates by thermal evaporation (for preparation conditions
see\cite{DeRose}), NbSe$_{2}$ cleaved in the vacuum chamber and a
thermally evaporated Nb/Pd sandwich sample deposited on a Si
substrate covered by a $500$ \AA\, SiO buffer layer. In the last
sample, a thick Nb layer (400 \AA) was covered by a thin Pd ($50$
\AA) layer to protect the Nb from oxidizing and to allow a
proximity induced gap with an amplitude close to that of bulk Nb
to appear in the Pd layer.

\section{Experimental Results}
\label{Experimental} In order to check the spatial resolution of
our Nb tip, we made topographic scans of an in-situ cleaved
NbSe$_{2}$ sample. A current image scan of 165 \AA\, $\times$ 165\AA\,
measured at a temperature of $4.2\,$ K is shown in Figure 2 which
clearly shows the atomic lattice of the surface. This confirms
that indeed the Nb tip is atomically sharp.

\begin{figure}
\centering
  \includegraphics[angle=-90, width=6cm]{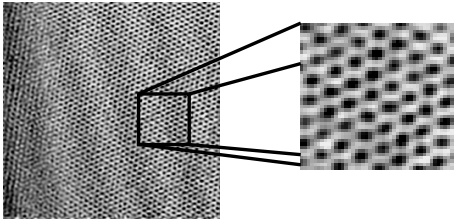}\\
  \caption{165 \AA\, $\times$ 165\AA\ current image of NbSe$_{2}$ sample obtained at a temperature of
  $4.2$K with an Nb tip showing atomic resolution. Inset shows a magnification of an area
  of $\sim$ 40\AA\,$\times$40\AA which shows clearly a hexagonal structure of the lattice}
\label{nbse2}
\end{figure}

\begin{figure}
\centering
  \includegraphics[width=6cm]{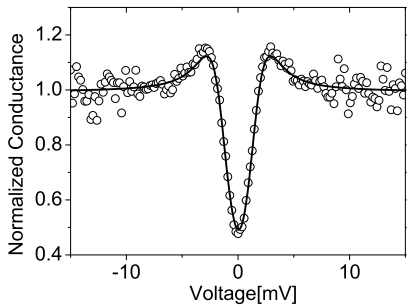}\\
  \caption{Conductance spectra of an Nb tip/ Au sample vacuum tunneling
  junction with BCS fit. $\Delta=$1.47meV,$\Gamma$ =0.34 meV, T=5.8K.}\label{temp}
\end{figure}

%%%%%%%%%%%%%

%%%%%%%%%%%%%%%%%%%%%

Figure 3 shows a typical NIS conductance curve measured using an
Nb tip and an Au sample at a temperature of $5.8$ K. The
conductance spectra were obtained by numerically differentiating
the measured I,V curves which were obtained by applying a digital
voltage ramp and measuring the current. In general the spectra are
symmetric with respect to zero bias representing the symmetry of
the BCS DOS with respect to energy. The depression in the
conductance near zero voltage due to the SC gap is clearly visible
and is followed, at higher voltage, by clear peaks as expected for
the BCS DOS. The additional oscillations visible at even higher
voltages are not symmetric with respect to zero voltage and are
due to experimental noise. The measured spectrum is fitted using
the BCS DOS with an addition of a life-time broadening parameter
\cite{Dynes}. The best fit is given using a gap value of $1.47$
meV which matches the value measured for bulk Nb and a broadening
parameter $\Gamma=0.34$ meV. The broadening parameter accounts for
both experimental broadening, resulting from bias jitter, and
inelastic scattering or pair breaking processes.

\begin{figure*}[t]
\centering
  \includegraphics{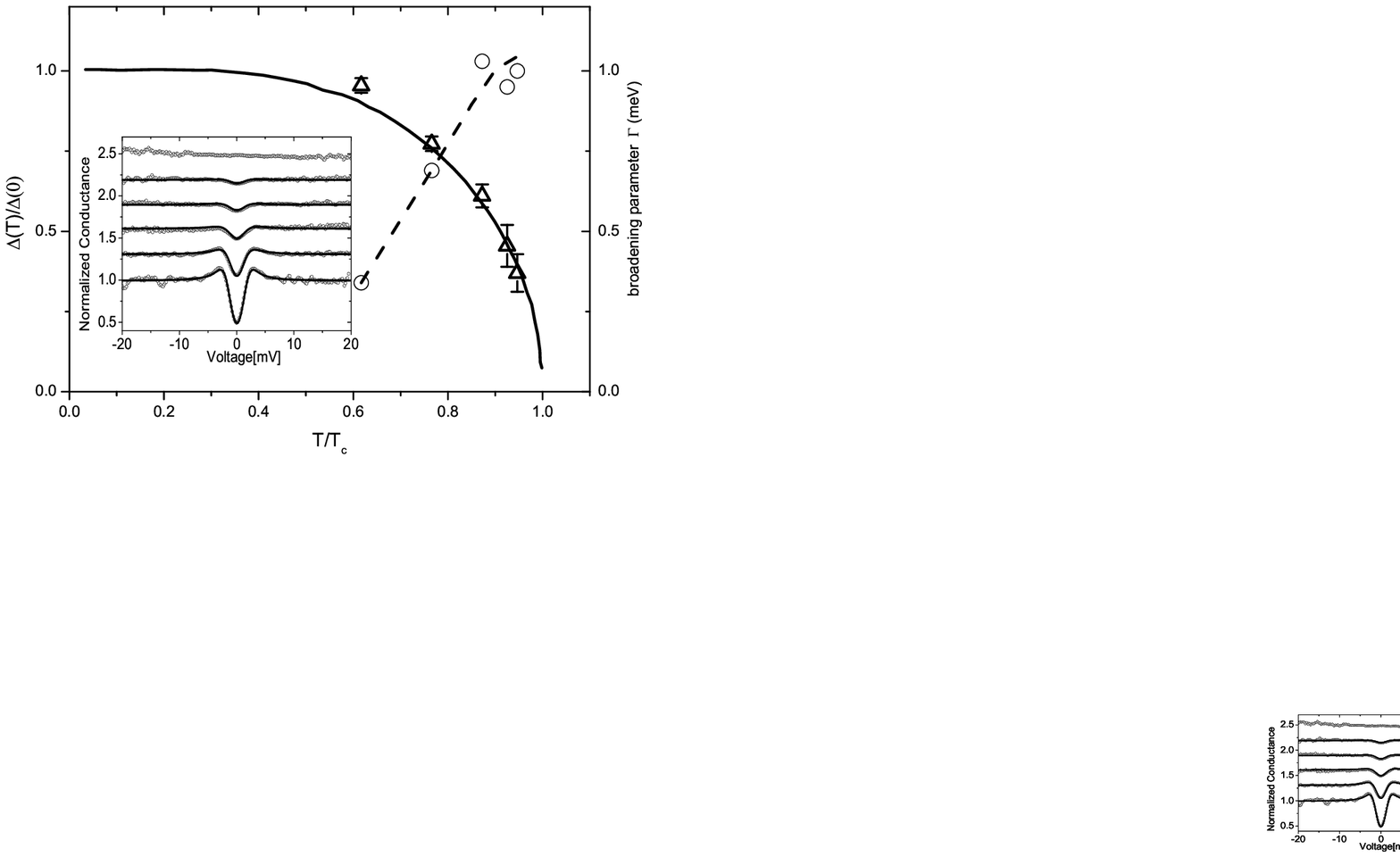}\\
  \caption{Values of the BCS fit parameters, the order parameter $\Delta/\Delta$
  (T=0)(triangles) and the broadening parameter $\Gamma$ (circles) as a function of $T/T_{c}$ for an Nb tip/ Au sample vacuum tunnelling junction.
$\Delta(0)$=1.53 meV and $T_{c}$=9.3K. The solid line gives the
theoretical prediction calculated according to the BCS theory.
Dashed line is a guide to the eye.
  Inset shows the measured conductance spectra at temperatures of 5.8K, 7.2K, 8.2K, 8.7K, 8.9K and 9.5K and the theoretical fit.
  The conductance spectra and the fits are shifted for clarity.}\label{tempfit}
\end{figure*}

We find the gap value to be reproducible in different positions on
the same sample and using several Nb tips and Au samples, while
the value of the broadening parameter changes between $0.34-1$
meV.

Figure 4 shows the values of the BCS fit parameters, the
normalized order parameter $\Delta/\Delta(T=0)$ and the life-time
broadening parameter $\Gamma$ as a function of the normalized
temperature, $T/T_{c}$, where we have used $\Delta(0)=1.53$  meV
and $T_{c}=9.3\,$K. The solid line depicts the theoretical
temperature dependence according to the BCS theory and is in
agreement with the values extracted from the experimentally
measured spectra. The inset shows the measured conductance spectra
and the theoretical fit in the range of 5.8 K to 9.5 K. One
clearly observes the closing of the SC gap as the temperature is
increased. At $8.9$ K a gap is still visible while at $9.5$ K a
flat conductance curve is measured, indicating a transition of the
tip to the normal state at $9.2\pm0.3$ K. This temperature is in
agreement with the one measured for bulk Nb samples, $T_{c}=9.2$ K
and with the value used for creating the normalized temperature
curve ($T_{c}=9.3$ K). The life-time broadening parameter is found
to increase with temperature (circles in fig 4), starting from
0.34 meV at T=5.8K and saturating at around 1meV as we approach
$T_{c}$. Naaman et al. \cite{Naaman} reported on an increase in
the broadening parameter for their Pb tips with $\Gamma=0.16$ meV
at 2 K and rising up to $\Gamma=0.4$ meV at T=6 K. As our
pre-amplifier is located on the microscope, and thus is subject to
roughly the same temperature changes as the junction, we expect
our bias jitter to increase with temperature due to increased
fluctuations in the amplifier's virtual ground. Therefore it is
impossible at this stage to determine if the increase in the value
of $\Gamma$ has also an intrinsic source, such as a rise in the
pair breaking rate near the tip apex as a function of temperature.

\begin{figure}[h]
\centering
  \includegraphics[width=8cm]{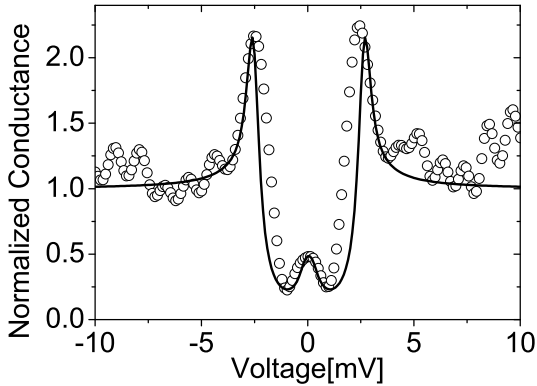}\\
  \caption{Conductance spectra of a Nb tip/ NbPd sample vacuum
  tunneling junction fitted with  SIS calculated curve using the SC
  BCS DOS for both electrodes.
  Fit parameters $\Delta_{Nb}$=1.4 meV, $\Delta_{NbPd}$=1.2meV,
  $\Gamma_{1}$=0.24 meV,$\Gamma_{2}$=0.12 meV,T=6.6K}
  \label{sis}
\end{figure}

In order to verify that indeed the use of a SC tip enhances the
spectral features measured on a SC sample we have used an Nb/Pd
(2000\,\AA/50\,\AA) sandwich type sample, in which
superconductivity is induced in the thin Pd layer by the thick Nb
layer due to the proximity effect. The covering Pd layer protects
the Nb layer from oxidizing which is necessary since the sample
was grown in a different vacuum system than the one used for the
STM measurements.  Figure 5 shows the spectra measured at a
temperature of 6.6K. At this temperature, a zero-bias peak in the
conductance is clearly observed. This peak is typical for SIS
junctions at finite temperature and is the result of the enhanced
tunnelling probability of thermally excited quasi particles.
Additionally, the amplitude of the gap-edge peaks is considerably
enhanced in comparison to the NIS case (fig 3) and the normalized
conductance value reaches a minimum value of $0.25$ much lower
than the one obtained for the NIS case $0.5$. The oscillations
observed at voltages above the gap are due to experimental noise.
The best fit of the data is shown in the figure (solid line) and
the fit parameters ($\Delta_{1}=1.4$ meV, $\Delta_{2}=1.2$ meV)
confirm that indeed both electrodes are SC. The reduced gap value
in Nb/Pd in comparison to that of the Nb tip is reasonable as it
is induced by the proximity effect while the Nb gap for the tip
fits the value expected at T=6.6K.

\section{Magnetic Field}
\label{Magnetic}
\begin{figure*}
\centering
  \includegraphics{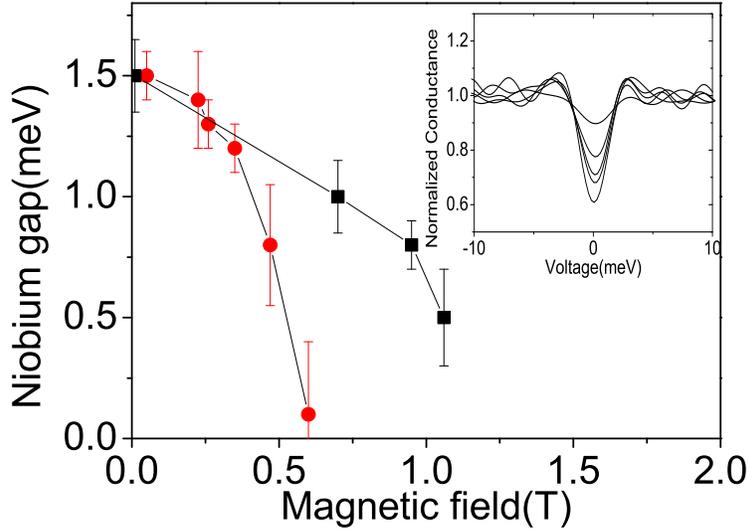}\\
  \caption{Field dependence of the Nb order parameter extracted by
  a BCS fit of Nb tip Au sample tunneling conductance spectra for
  two different Nb tips measured at T=5.5K. Inset shows the conductance
  spectra from which gap values marked by circles were extracted in different
  magnetic fields: 0.05, 0.24, 0.3, 0.35, 0.47 T}\label{magnet}
\end{figure*}

In order to check the magnetic field response of our Nb tip we
have studied the NIS configuration using an Nb tip and an Au
sample. We then applied a magnetic field parallel to the Nb wire
whose apex forms the tip. Starting from zero field we have
measured the tunneling conductance as a function of field up to
1.5 T. A typical result is shown in the inset of figure 6. The
data shown is Fourier filtered to remove the high frequency noise
appearing in our system due to the effect of the magnetic field on
the pre-amplifier. As the field was increased we observed a
closing of the gap, however even at fields higher than the second
critical field of bulk Nb ($H_{c2}=0.21$ T at $T=0$ K) a non-zero
gap could be clearly observed. We note the lowest field in which
we found no evidence for a gap in the spectra as H$_{ctip}$. The
values of H$_{ctip}$, all obtained at $T\sim5$ K, varied from one
tip to the other in the range of $0.5$ T to $1$ T, but were always
considerably enhanced in comparison to the value measured in bulk
samples. As our measurements were made at $T\sim5$ K, the actual
values of the critical field at zero temperature should be even
higher.

The enhancement of the critical
field is clearly seen in figure 6 where we present the value of
the SC order parameter as a function of magnetic field, obtained
by fitting the measured curves using the BCS formula, for two of
our tips. We consider this enhancement to be a result of the
special geometry of the tip. Far from its apex the tip has the
properties of a bulk sample thus becoming normal when the applied
field exceeds the value of the bulk one. However as the tip
narrows towards its apex the mean free path and consequently the
effective coherence length become limited and are reduced in
comparison to their bulk values, thus an enhancement of the
critical field similar to that calculated for thin films in the
dirty limit \cite{Tinkham} is expected. We note that in such a
case the field dependence of the gap is expected to be quadratic,
in agreement with our experimental results. Misko et al. have
theoretically studied the field dependence of a superconductor
with a multi conoid geometry by solving the Ginzburg Landau
equations \cite{Misko}. Indeed they have found an enhancement of
$H_{c2}$ towards the apex of the cone. The model was solved
numerically for the case of Pb, taking $\lambda_{Pb}=39$ nm and
$\xi_{effective}=10$ nm (set by the width of the bridge) and showed
an enhancement of $H_{c2}$ by a factor of $\sim5$. Experimentally
this enhancement was observed in a Pb nano-bridge formed between
an STM tip an a Pb sample by Rodrigo et al.\cite{Rodrigo} The
predicted enhancement fits well with our maximal measured value of
$H_{ctip}\sim1$ T. The results should be applicable also to Nb as
$\lambda_{Nb}=\lambda_{Pb}\sim39$ nm and the difference in
coherence length ($\xi_{Nb(bulk)}=38$ nm, $\xi_{Pb(bulk)}=83$ nm)
becomes irrelevant as the width of the confinement region is much
smaller than both.

Within the above model one can account for the
different values of $H_{ctip}$ found experimentally in different
tips as a result of variations in the tips geometries.
Moreover, the zero field spectra of the two
tips are almost identical in contrast to their field dependencies,
as indicated in fig.6, which exhibit $H_{ctip}$ values differing by at least $70$
percent.
This perfectly fits with the model,
since the zero field conductance should not be effected by the tip geometry.
\section{Conclusions}
\label{con} To conclude we have developed a simple method for the
fabrication of superconducting Nb tips suitable for atomic
resolution STM. The tunneling spectra show a gap value and a
critical temperature matching the values found in bulk Nb samples
but a highly enhanced value of the critical field, up to 5 times
larger.  The tips exhibit the expected enhancement of the STM
spectroscopic resolution in the SIS configuration. Future
prospects could include measurements of Josephson currents as a
function of position, i.e. a Josephson STM. Using the theoretical
condition for the measurement of Josephson currents,
$R<\Delta(T=0)$$R_{0}/(2k_{B}T)$ \cite{Smakov}, where
$R_{0}=\hbar/e^{2}\sim4 k\Omega$ is the quantum resistance, yields
for Nb : $R<36K\Omega/(T/1[K])$, which at this point could not be
obtained due to limitations in our present experimental setup.
\section{Acknowledgements}
The authors would like to thank Tristan Cren for fruitful
discussions. One of the authors, A.K. would like to thank R.B.
Beck and O. Naamam for their advice.

\end{document}